\documentclass[twocolumn,pre,showpacs]{revtex4}
\usepackage{epsfig}
\usepackage{graphicx}
\usepackage{psfig,epsfig}
\usepackage{dcolumn}
\usepackage{bm}

\def\e{\mbox{e}}

\begin{document}

\title{Jamming transition in granular media: A mean field approximation and
numerical simulations}
\author{A. Fierro $^{a,b}$, M. Nicodemi $^{a,b}$,
M. Tarzia $^{a}$, A. de Candia $^{a}$, A. Coniglio $^{a,b}$}

\affiliation{${}^a$ Dipartimento di Scienze Fisiche, Universit\`{a} degli Studi
di Napoli ``Federico II'', INFM and INFN, via Cinthia, 80126 Napoli, Italy}

\affiliation{${}^b$ INFM - Coherentia, Napoli, Italy}
\date{\today}

\begin{abstract}

In order to study analytically the nature of the jamming
transition in granular material, we have considered a cavity
method mean field theory, in the framework of a statistical
mechanics approach, based on Edwards' original idea. For simplicity
we have applied the theory to a lattice model and a transition
with exactly the same nature of the glass transition in mean field
models for usual glass formers is found. The model is also
simulated in three dimensions under tap dynamics and a jamming
transition with glassy features is observed. In particular two
step decays appear in the relaxation functions and dynamic
heterogeneities resembling ones usually observed in glassy
systems. These results confirm early speculations about the
connection between the jamming transition in granular media and
the glass transition in usual glass formers, giving moreover a
precise interpretation of its nature.


\end{abstract}

\date{\today}

\maketitle

\section{Introduction}
A deep connection between glass
transition in molecular glass formers,  structural arrest in
colloidal systems, and  jamming transition in granular media
\cite{coniglio,LN,OHern1,Danna,Mehta,OHern2} has often been stressed in the
past few years. In spite of the fact that these systems are very
different one from each other, varying suitably the control
parameters, a slowdown and a subsequent structural arrest in a
solid-like disordered state are found in each of them. In
\cite{LN,OHern2} a possible phase diagram for jamming is
suggested, which takes into account the fact that jamming is
obtained either raising the volume fraction or lowering the
temperature or lowering the applied stress. Colloidal suspensions
and molecular glass formers are both thermal systems,
and it is commonly accepted that both colloidal glass transition
and molecular glass transition are of the same type despite of the
fact that  different control parameters may drive the transition.
The case of granular materials is instead very different: They are
athermal systems, since the thermal fluctuations are significantly
less than the gravitational energy and the system cannot explore
the phase space without any external driving. Nevertheless an
exceedingly slowing down is observed when a granular material is
shaken at low shaking amplitude, or flows under a low shear
stress, with strong analogies with the slowing down observed in
glass formers. Experimental and numerical studies
\cite{Danna,OHern2,NCH,Mehta} have confirmed this connection,
however its precise nature is still unclear \cite{OHern1,OHern2}.

In the present paper in order to study this connection we apply a
statistical mechanics approach to granular media. This approach,
which has been extensively developed in previous works \cite{e1, fnc}, is
based on an elaboration of the original ideas suggested by
Edwards \cite{Edwards}. The basic assumption is that for a granular
system subject to an external drive (e. g. tapping), after having
reached stationarity, time averages coincide with suitable
ensemble averages over the ``mechanically stable" states. We have
shown \cite{fnc} that this assumption works for different lattice
models namely that a generalized Gibbs distribution of the stable
states describes with good approximation the stationary state
attained by the system under tapping dynamics. Here each tap
consists in raising the bath temperature to a finite value (called
tap amplitude) and, after a lapse of time (called tap duration)
quenching the bath temperature back to zero. By cyclically
repeating the process  the system explores the space of the
mechanically stable states.

We thus consider one of the above lattice model for which the
statistical mechanics approach works. The model is made up of hard
spheres under gravity. Then we apply standard statistical
mechanics methods in order to investigate analytically the
existence and the nature of a possible jamming transition. More
precisely we consider the Bethe-Peierls approximation using the
cavity method \cite{MP,Biroli}: By changing the control parameter
a phase transition from a fluid to a crystal is found, and, when
crystallization is avoided, a glassy phase appears. The nature of
this glassy phase is analogous to that found in mean field models
for glass formers \cite{Kurchan, Biroli, PC03}: In particular we
observe a dynamical transition, where an exponentially high number
of metastable states appears, and at a lower temperature a
thermodynamic discontinuous phase transition to a glassy state. A
brief account of these calculations was given in a previous
Letter \cite{epl}. We also studied \cite{epl} the model in
$3d$ by means of numerical simulations, and we found that the
model under taps has a transition from a fluid to a crystal, in a
very good agreement with he mean field approximation. However the
numerical simulation was not suitable to study the glass
transition since the model showed a strong tendency towards
crystallization.

For this reason we study here a variant of the model \cite{PC03}
which has the virtue of avoiding crystallization. We find that the
system under gravity evolved by Monte Carlo taps presents features
characteristic of real granular media \cite{Bideau, Clement}, and
at low tap amplitudes a dynamical transition with properties
recalling those of usual glass formers. In particular we observe a
dynamical non linear susceptibility with a maximum at increasing
time: This behavior, typical of glass formers, is usually
interpreted as the sign of dynamic heterogeneities in the system.

In conclusions the results confirm early
speculations about the deep connection between the jamming
transition in granular
media and the glass transition in usual glass formers, giving moreover a
precise interpretation to its nature.



In Sect. \ref{meanfield} the mean field phase diagram is discussed. The details
of calculations are presented in App.s \ref{app1} and \ref{app2}. In particular
in App. \ref{app2} the self-consistency equations obtained using the cavity
method are shown.  In Sect. \ref{hardsphere} the $3d$ model is presented and the
numerical results are shown.

\section{Mean field solution in the Bethe-Peierls approximation}
\label{meanfield}
The model is a monodisperse hard sphere system (with diameter $\sqrt{2} a_0$)
under gravity, constrained to move
on the sites of a cubic lattice of spacing $a_0=1$. The Hamiltonian is given by:
\begin{equation}
{\cal H} = {\cal H}_{HC} + mg \sum_i n_i  z_i
\label{H1}
\end{equation}
where $z_i$ is the height of site $i$, $g=1$ is the gravity acceleration,
$m=1$ the grain mass, $n_i\in\{0,1\}$ is the occupancy variable
(absence or presence of a grain on site $i$)
and ${\cal H}_{HC}(\{n_i\})$ is the hard core term
preventing two nearest neighbor sites being simultaneously occupied.

We have shown in previous papers \cite{fnc} that
the model, Eq.~(\ref{H1}), evolving by means of a tap dynamics can be described
in good approximation by
a generalized Gibbs distribution of the ``mechanically stable'' states (i.e.
the states where the system is found at rest).  In particular the weight of a
given state, $\{n_i\}$, is:
\begin{equation}
\e^{-\beta H(\{n_i\})}{\cdot} \Pi(\{n_i\}),
\end{equation}
where $T_{conf} = \beta^{-1}$ is a thermodynamic parameter,
called ``configurational temperature'', characterizing the distribution.
The operator $\Pi(\{n_i\})$ selects mechanically stable states:
$\Pi(\{n_i\})=1$ if $\{n_i\}$ is ``stable'', or else $\Pi(\{n_i\})=0$.
The system partition function is thus the following \cite{fnc}:\begin{equation}
{\cal Z} =\sum_{\{n_i\}} \e^{-\beta H(\{n_i\})}{\cdot} \Pi(\{n_i\})
\label{Z}
\end{equation}
where the sum runs over all microstates, $\{n_i\}$.

\begin{figure}[ht]
\centerline{
\psfig{figure=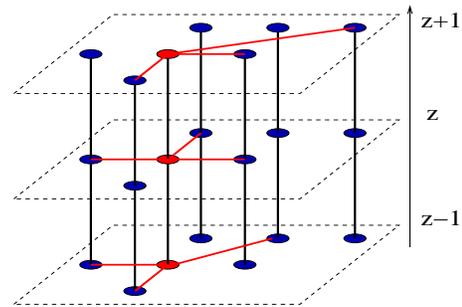,width=6cm,height=4cm,angle=0}
}
\vspace{-.3cm}
\caption{In the mean field approximation,
the grains are located on a Bethe lattice, sketched in the figure,
where each horizontal layer is a random graph of given connectivity.
Homologous sites on neighboring layers are also linked and the overall
connectivity, $c$, of the vertices is $c\equiv k+1=5$.}
\label{Blattice}
\end{figure}

In the present section we show the phase diagram of the model, Eq.~(\ref{H1}),
obtained using a mean field theory in the Bethe-Peierls approximation (see
\cite{MP, Biroli} and ref.s therein),
based on a random graph (plotted in Fig.~\ref{Blattice})
which keeps into account that the gravity breaks up the symmetry along the $z$
axis. This lattice is made up by $H$ horizontal layers
(i.e., $z\in\{1,...,H\}$).
Each layer is a random graph of connectivity, $k-1=3$. Each site in layer
$z$ is also
connected to its homologous site in $z-1$ and $z+1$
(the total connectivity is thus $k+1$).
Locally the graph has a tree-like structure but there
are loops whose length is of order $\ln N$, insuring geometric frustration.
In the thermodynamic limit only very long loops are present.
The details of calculations are given in appendices~\ref{app1} and~\ref{app2}
(see also Ref.s~\cite{epl, prl} where this mean field theory was
first introduced).

We solve the recurrence equations found in the Bethe-Peierls approximation in
three cases: 1) A fluid-like homogeneous phase; 2) a crystalline-like
phase characterized by the breakdown of the horizontal translational
invariance; 3)  a glassy phase described by a $1$-step Replica Symmetry
Breaking (1RSB).  The details of the calculations are shown in Appendices.

\begin{figure}[ht]
\vspace{-0.5cm}
\begin{center}
\mbox{\epsfysize=8cm\epsfbox{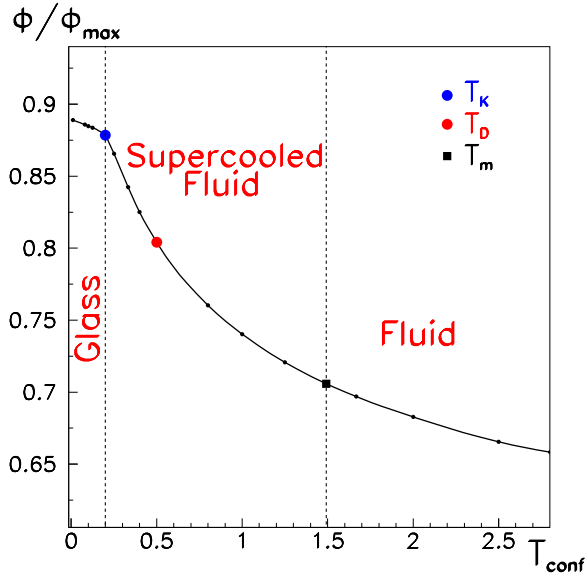}}
\end{center}
\caption{ The density, $\Phi\equiv N_s/(2\langle z \rangle-1)$,
for $N_s=0.6$ as a function of $T_{conf}$. $\Phi_{max}$ is the maximum density
reached by the system in the crystal phase.}
\label{phi_T}
\end{figure}

The results of the calculations are summarized in Fig.~\ref{phi_T}, where the
bulk density at equilibrium, $\Phi\equiv N_s/(2\langle z\rangle-1)$
\cite{notadens}
(where $\langle z\rangle$ is the average height)
is plotted as a function of the configurational temperature, $T_{conf}$, for a
given value of the number of grains per unit surface, $N_s$.
We found that at high $T_{conf}$
a homogeneous solution corresponding to the fluid-like phase is found.
By lowering $T_{conf}$ at $T_m$
a phase transition to a crystal phase (an anti-ferromagnetic solution with a
breakdown of the translation invariance) occurs.
The fluid phase still exist below $T_m$ as a metastable phase
corresponding to a supercooled fluid when crystallization
is avoided. Finally a 1RSB solution (found with the cavity
method \cite{MP}), characterized by the presence of a large number of
local minima in the free energy \cite{MP}, appears
at $T_D$, and becomes stable at a lower point $T_K$, where
a thermodynamic transition from the supercooled fluid
to a 1RSB glassy phase takes place.
The temperature $T_D$, which is interpreted in mean field
as the location of a  dynamical transition where the relaxation time diverges,
in real systems might instead correspond to a crossover in the dynamics
(see \cite{Kurchan, Biroli, Toninelli} and Ref.s therein).
$\Phi(T_{conf})$ has a shape very similar to that observed in the
``reversible regime'' of tap experiments \cite{Knight,Bideau}.
The location of the glass transition, $T_K$, corresponds
to a cusp in the function $\Phi(T_{conf})$. The dynamical crossover point $T_D$
might  correspond to the position of a characteristic shaking amplitude
$\Gamma^*$ found in experiments and simulations where the ``irreversible''
and ``reversible'' regimes approximately meet.

\begin{figure}[ht]
\centerline{
\psfig{figure=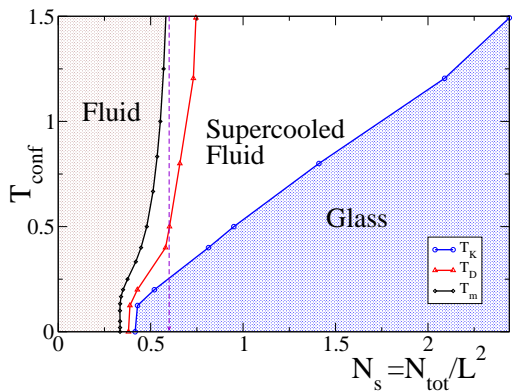,width=6cm,angle=-90}}
\caption{The system mean field phase diagram is plotted in the plane
of its two control parameters $(T_{conf},N_s)$.}
\label{MFPD}
\end{figure}

In Fig.~\ref{MFPD} the phase diagram obtained by varying $N_s$ is shown. The
dashed vertical line in figure corresponds to the value of $N_s$ chosen in
Fig.~\ref{phi_T}.

The model, Eq.~(\ref{H1}), simulated in $3d$ by means of Monte Carlo tap
dynamics \cite{epl} presents a transition from a fluid  to a crystal
as predicted by the mean field approximation, density profiles in good
agreement with the mean field ones, and in the fluid phase a large increase of
the relaxation time as a function of the inverse tap amplitude.
In the following section we study a more complex model for hard spheres,
where an internal degree of freedom allows to
avoid crystallization \cite{PC03}.

\section{Hard spheres with an internal degree of freedom}
\label{hardsphere}
The Hamiltonian of the model is
\begin{equation}
{\cal H}  =\sum_{\langle ij \rangle} n_i n_j \phi_{ij}(\sigma_i,\sigma_j)
+ mg \sum_i n_i  z_i,
\label{H2}
\end{equation}
where $z_i$ is the height of site $i$, $g=1$ is the gravity acceleration,
$m=1$ the grain mass, $n_i\in\{0,1\}$ is the occupancy variable (absence
or presence of a grain on site $i$), $\sigma_i\in\{1,\ldots,q\}$ represents the
internal degree of freedom  (which we call spin), and
$\phi_{ij}(\sigma_i,\sigma_j)$ is the interaction energy between spins.
Different values of the spin correspond to different positions of the
particle inside the cell.  It is reasonable that a few number
of internal states might be enough to catch the main features
of real systems.

\begin{figure}[ht!]
\hspace{-1cm}\centerline{\mbox{\epsfysize=6cm\epsfbox{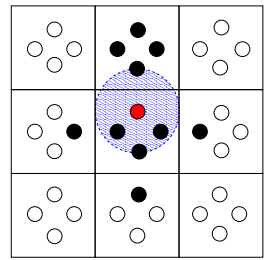}}}
\vspace{-2.5cm}
\caption{The model in two dimensions: the space is partitioned in square cells,
and each cell can be occupied by at most one particle
in anyone of the four shown positions (little circles). A particle in
any given position (large shaded circle) excludes the presence of
particles in any of the black colored positions.}
\label{fig-model}
\end{figure}

As in Ref.~\cite{PC03} we study a  simple realization of the model
described by Eq.~(\ref{H2}). Interpreting the spin as position of the particle
in the cell, our choice can be easily visualized in $2d$, as shown in Fig.~
\ref{fig-model}. We
partition the space in square cells, and subdivide each cell into four
internal positions (namely $q=4$). When a cell is occupied by a particle in any
given position, a hard-core repulsion excludes the presence of
particles in some of the internal states of the neighboring
cells (namely the interaction $\phi_{ij}(\sigma_i, \sigma_j)$ is chosen zero
if the
positions $\sigma_i$ and $\sigma_j$ are ``compatible'', and infinite otherwise).
This choice can be interpreted as a coarse grained version of a
hard sphere system in the continuum.  In $3d$ we subdivide the
space into cubic cells, and considers six internal positions
instead of four.

\begin{figure}[ht]
\begin{center}
\mbox{\epsfysize=7cm\epsfbox{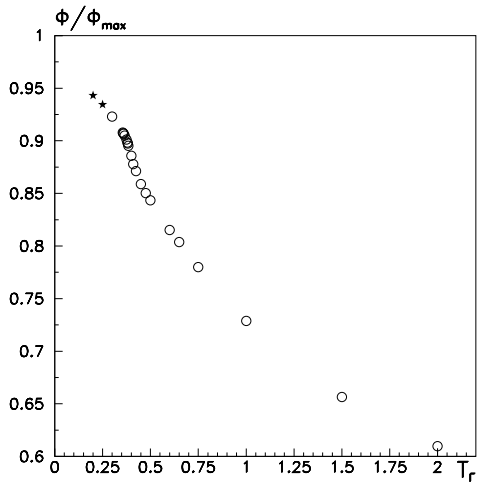}}
\end{center}
\vspace{-0.5cm}
\caption{The bulk density, $\Phi\equiv N/L^2(2\langle z\rangle-1)$,
is plotted as function of $T_\Gamma$ for $\tau_0=10~MCsteps/particle$. The empty
circles correspond to stationary states, and the black stars to out of
stationarity ones. $\Phi_{max}$ is the maximum density reached by the system in
the crystal phase, $\Phi_{max}=6/7$.}
\label{phi}
\end{figure}
\begin{figure}[ht]
\begin{center}
\mbox{\epsfysize=7cm\epsfbox{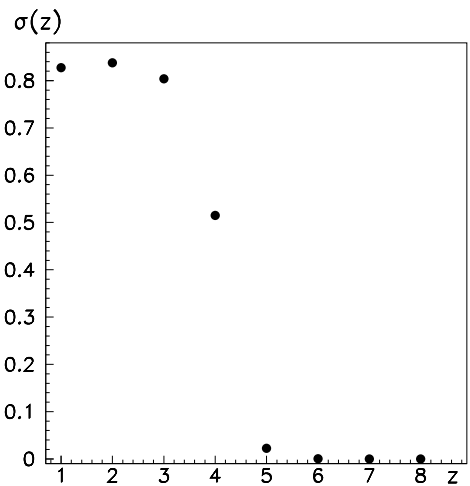}}
\end{center}
\vspace{-0.5cm}
\caption{The density profile, $\sigma (z)$, as function of the height, $z$, for
$T_\Gamma=~0.20$ and $\tau_0=10~MCsteps/particle$.}
\label{prof}
\end{figure}

In the Monte Carlo simulations, $N=~433$ grains are confined in a $3d$ box
of linear size $L=~12$ (i.e. $N_s=~3$), between hard walls in the vertical
direction and with periodic boundary
conditions in the horizontal directions.  We perform a standard Metropolis
algorithm on the system.  The particles,  initially prepared in a random
configuration, are subject to taps, each one followed by a
relaxation process. During a tap, for a time $\tau_0$ (called tap
duration), the temperature is set to the value $T_{\Gamma}$ (called tap
amplitude), so that particles have a finite probability, $p_{up}\sim
e^{-mg/T_{\Gamma}}$, to move upwards. During the relaxation the
temperature is set to zero, so that particles can only reduce the energy,
and therefore can move only downwards. The relaxation stops when the
system has reached a blocked state, where no grain can move downwards.
Our measurements are performed at this stage when
the shake is off and the system is at rest. The time, $t$,
is the number of taps applied to the system.

In the following the tap duration is fixed, $\tau_0=10 MCsteps/particle$, and
different tap amplitudes, $T_\Gamma$, are considered.
In Fig.~\ref{phi} the bulk density, $\Phi \equiv N/
L^2(2\langle z \rangle-1)$, is plotted as a function of $T_\Gamma$:
$\Phi(T_\Gamma)$ has a shape resembling that found in the
``reversible regime'' of tap experiments \cite{Knight, Bideau}, and moreover
very similar to that
obtained in the mean field calculations and shown in Fig.~\ref{phi_T}.
At low shaking amplitudes (corresponding to high bulk densities) a strong
growth of the equilibration time (i.e. the time necessary to reach stationarity)
is observed, and for the lowest values here considered (the black stars in
Fig.~\ref{phi}) the system remains out of stationarity. In this region the
density profile, $\sigma (z)\equiv 1/L^2\sum_i n_i(z)$ (where the sum $\sum_i$
is done over the sites $i$ in the layer $z=~1, \dots, L$ \cite{notapart}),
is almost constant until a given layer and sharply decays to zero (see Fig.
~\ref{prof}), as found in real granular media \cite{Clement}.
In conclusions the system here studied presents a jamming transition at low
tap amplitudes as found in real granular media.

In order to test the predictions of the mean field calculations,
in the following we measure quantities usually important in the
study of glass transition: The relaxation functions, the
relaxation time
and the dynamical susceptibility, connected to the presence a
dynamical correlation length.

\begin{figure}[ht]
\begin{center}
\mbox{\epsfysize=7cm\epsfbox{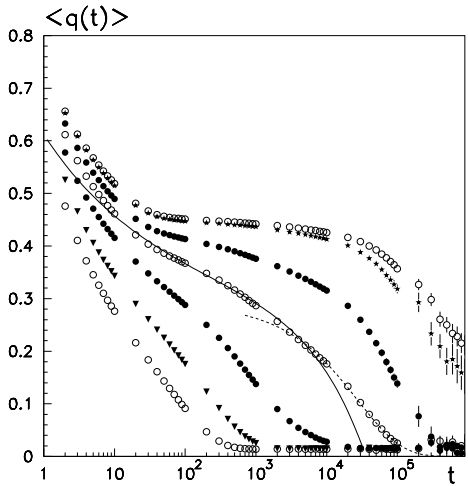}}
\end{center}
\vspace{-0.5cm}
\caption{The ``equilibrium'' autocorrelation function, $\langle q(t) \rangle$,
plotted as function
of $t$, for tap amplitudes $T_\Gamma=~0.60,~0.50,~0.425,
~0.40,~0.385,~0.365,~ 0.36$  (from bottom to top). The continuous line in
figure is the $\beta-$correlator of the mode coupling theory 
with exponent parameters
$a=~0.30$ and $b=~0.52$.  The dashed line is a stretched exponential $\propto
exp[-(t/\tau)^\beta]$ with $\beta=~0.70$.}
\label{tti}
\end{figure}

In particular we calculate the two-time autocorrelation
functions:
\begin{equation}
C(t,t_w)=\frac{1}{N}\sum_i\overline{ n_i(t)n_i(t_w) \vec {\sigma} _i(t){\cdot}
\vec{\sigma} _i(t_w)},
\end{equation}
where $\vec {\sigma} _i$ are unit length vectors, pointing
in one of the six coordinate directions, representing the position of the
particles inside the cell; the average $\overline{(\ldots)} $ is done over
$16-32$ different
realizations of the model obtained varying the random number generator in
the simulations, and the errors are calculated as the fluctuations
over this statistical ensemble.
For values of $t_w$ long enough, the system
reaches a stationary state, where the time translation invariance is
recovered, i.e., $C(t,t_w)=C(t-t_w)$.
In this time region, by averaging $C(t',t_w)$
over $t'$ and $t_w$ such that $t=t'-t_w$ is fixed, we calculate
the ``equilibrium'' autocorrelation functions
\begin{equation}
\langle q(t) \rangle= \langle C(t'-t_w) \rangle,
\end{equation}
and the dynamical non linear susceptibility
\begin{equation}
\chi(t)=\langle q(t)^2 \rangle - \langle q(t) \rangle^2.
\end{equation}
As shown in Fig.~\ref{tti}, at low values of the tap amplitudes, $T_\Gamma$,
two-step decays appear, well fitted in the intermediate time region,
by the $\beta-$correlator predicted by the mode coupling theory for 
supercooled liquids \cite{mct, notaMCT} (the continuous
curve in Fig.~\ref{tti}), and at long time by stretched exponentials (the dashed
curve in figure).
The relaxation time, $\tau$, is defined as $\langle q(\tau)
\rangle \sim 0.1$. 

\begin{figure}[ht]
\begin{center}
\mbox{\epsfysize=7cm\epsfbox{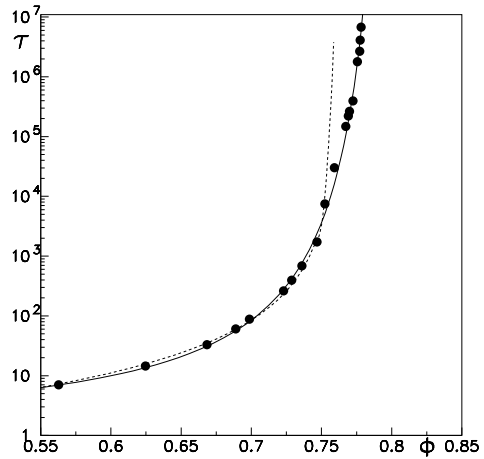}}
\end{center}
\vspace{-0.5cm}
\caption{The relaxation time, $\tau$, as function of the bulk
density, $\Phi$. The continuous line is a Vogel-Fulcher,
$e^{A/(\Phi_c-\Phi)}$,with $\Phi_c=0.81\pm 0.01$ and $A=0.49\pm 0.10$. The 
dashed line is a power law, $(\Phi_D-\Phi)^{-\gamma_1}$, with
$\Phi_D=0.76\pm 0.01$ and $\gamma_1=2.04\pm 0.10$.}
\label{tau_phi}
\end{figure}

In Fig.~\ref{tau_phi} the relaxation time, $\tau$, is plotted as a function of 
the density, $\Phi$. As found in many glass forming liquids, $\tau(\Phi)$ is 
well fitted by a Vogel-Fulcher for the entire range, even if we can identify a 
first region where $\tau(\Phi)$ is fitted with good approximation by a power 
law. The power law divergence can be interpreted as a mean field behavior, 
followed by a hopping regime.
Note that the model, Eq.~(\ref{H2}), studied in absence of gravity
by means the usual Monte Carlo Metropolis \cite{PC03}, exhibits a
divergence of the relaxation time as a power law, and no crossover
to a hopping regime is observed. We suggest that in the present
case the tap dynamics favors the equilibration
via hopping precesses.

\begin{figure}[ht]
\begin{center}
\mbox{\epsfysize=7cm\epsfbox{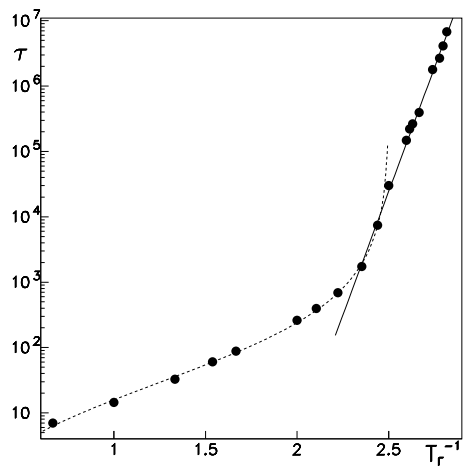}}
\end{center}
\vspace{-0.5cm} \caption{The relaxation time, $\tau$, as function
of the tap amplitude inverse, $T_\Gamma^{-1}$. The dashed line
is a power law, $(T_\Gamma-T_D)^{-\gamma_2}$, with
$T_D=0.40\pm 0.01$ and $\gamma_2=1.52\pm 0.10$. The continuous line
is an Arrhenius fit, $e^{A/T_\Gamma}$, with $A=17.4\pm 0.5$ (the data in this 
region are also well fitted by both a super-Arrhenius and
Vogel-Fulcher laws).}
\label{figure3}
\end{figure}
In Fig.~\ref{figure3} the relaxation time, $\tau$, is plotted as a function of 
the tap amplitude, $T_\Gamma$: A clear crossover
from a power law to a different regime is again observed around a
tap amplitude $T_D$, corresponding to the value of the density, $\Phi(T_D)
\simeq \Phi_D$, where a similar crossover has been found in Fig.~\ref{tau_phi}.

The divergence of the relaxation time at vanishing tap amplitude
is consistent with the experimental data of Philippe
and Bideau \cite{Bideau} and D'Anna {\em et al.} \cite{Danna}. 
Their findings are in fact consistent with an Arrhenius
behavior as function of the experimental tap
amplitude intensity. However a direct comparison with our data is not
possible since we do not know the relation between the
experimental tap amplitude and the tap amplitude in our simulations.
A more direct comparison would be possible if the experimental data 
were plotted as function of the bulk density, as we did in Fig.~\ref{tau_phi}.

\begin{figure}[ht]
\begin{center}
\mbox{\epsfysize=7cm\epsfbox{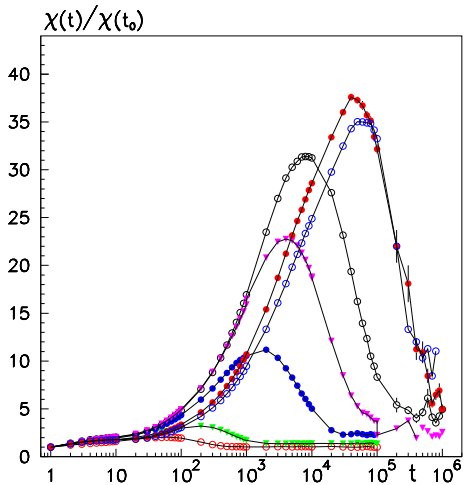}}
\end{center}
\vspace{-0.5cm}
\caption{The dynamical non linear susceptibility, $\chi(t)$, (normalized by
$\chi(t_0)$, the value at $t_0=1$) as a function of $t$, for tap amplitudes
$T_\Gamma=
0.60,~0.50,~0.425, ~0.41, ~0.40,~0.385, 0.3825$  (from left to right).}
\label{chi}
\end{figure}

The dynamical non linear susceptibility, $\chi(t)$, plotted in
Fig.~\ref{chi} at different $T_\Gamma$, exhibits a maximum at a
time, $t^*(T_\Gamma)$. The presence of a maximum in the  dynamical non linear
susceptibility is typical of glassy systems \cite{franz, glotzer}.
In particular the value of the maximum, $\chi(t^*)$, diverges in
the $p$-spin model \cite{franz} as the dynamical transition is
approached from above, signaling the presence of a diverging
dynamical correlation length.
In the present case the value of the maximum increases as $T_\Gamma$
decreases (except at very low $T_\Gamma$ where the maximum seems to
decrease \cite{nota_max}). The growth of $\chi(t^*)$ in our model 
suggests the presence of a growing
dynamical length also in granular media.


\section{Conclusions}

In conclusions using standard methods of statistical mechanics we
have investigated  the jamming transition in a model for granular
media. We have shown a deep connection  between the jamming
transition in granular media and the glass transition in usual
glass formers. As in usual glass formers the mean field
calculations obtained using a statistical mechanics approach to
granular media predict a dynamical transition at a finite
temperature, $T_D$, and, at a  lower temperature, $T_K$, a
thermodynamics discontinuous phase transition to a glass phase. In
finite dimensions 1) the dynamical transition becomes only a
dynamical crossover as also found in usual glass formers
\cite{Kurchan, Biroli,Toninelli} (here the relaxation time,
$\tau$, as a function of both the density and the tap amplitude,  
presents a crossover
from a power law to a different regime); and 2) the thermodynamics
transition temperature, $T_K$, seems to go to zero (the relaxation
time, $\tau$, seems to diverge only at $T_\Gamma\simeq 0$, even if a
very low value of the transition temperature is consistent with
the data).

\begin{acknowledgments}
We would like to thank M. Pica Ciamarra for many interesting discussions and 
suggestions. Work supported by EU Network Number  MRTN-CT-2003-504712, 
MIUR-PRIN 2002, MIUR-FIRB 2002, CRdC-AMRA, INFM-PCI.
\end{acknowledgments}

\appendix
\section{Mean field solution}
\label{app1}

We consider the Hamiltonian, Eq. (\ref{H1}), plus a chemical potential
term which controls the overall density. Hard Core repulsion prevents
two connected sites to be occupied at the same time.
We adopt a simple definition of ``mechanical
stability'': a grain is ``stable'' if it has a grain underneath.
For a given grain configuration $r=\{n_i\}$, the operator $\Pi_r$ has a
simple expression: $\Pi_r =\lim_{K\rightarrow\infty}\exp\left\{-K
{\cal H}_{Edw}\right\}$, where ${\cal H}_{Edw}=
\sum_i \delta_{n_i(z),1}\delta_{n_i(z-1),0}\delta_{n_i(z-2),0}$ (for clarity, we
have shown the $z$ dependence in $n_i(z)$).

\begin{figure}
 \begin{center}
    \includegraphics[scale=0.27]{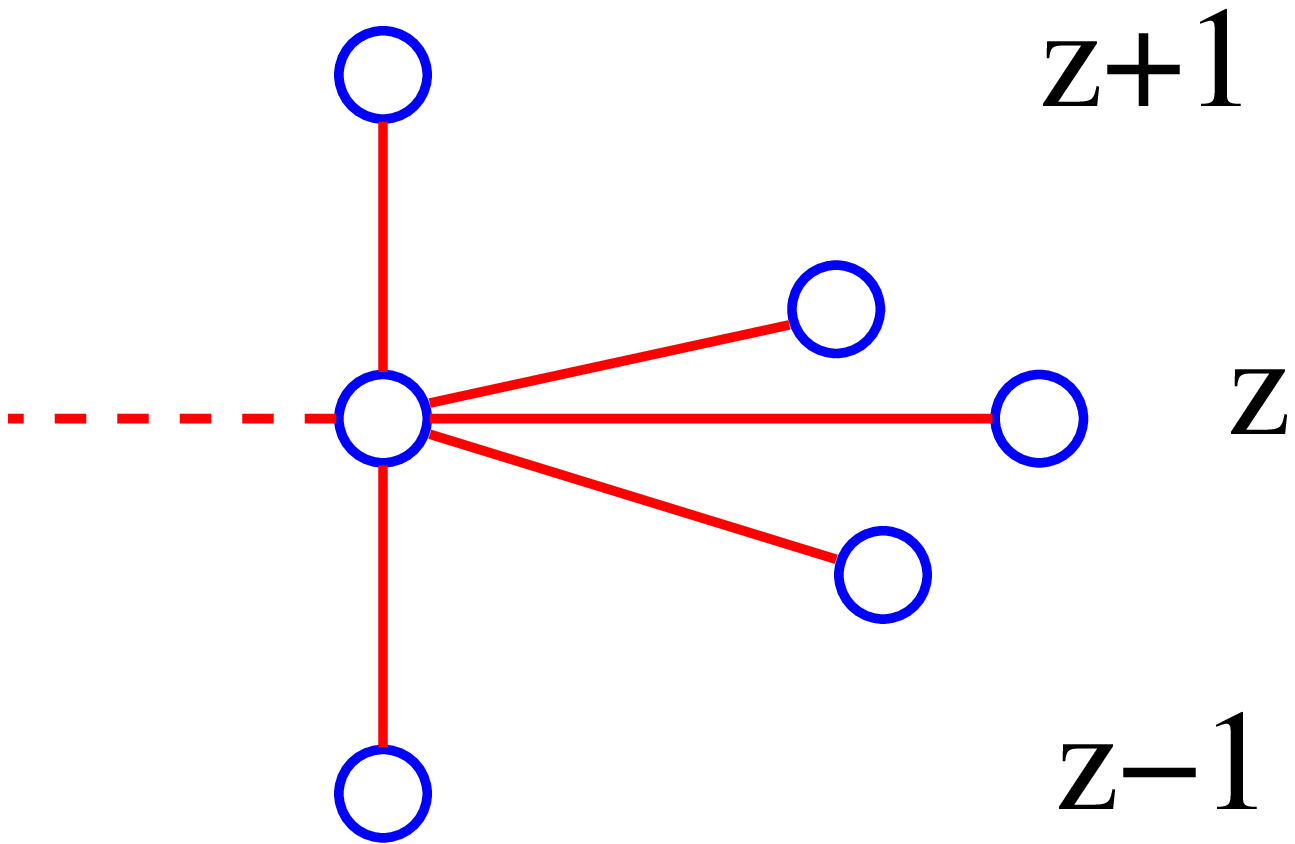} \mbox{~~~~}
\includegraphics[scale=0.27]{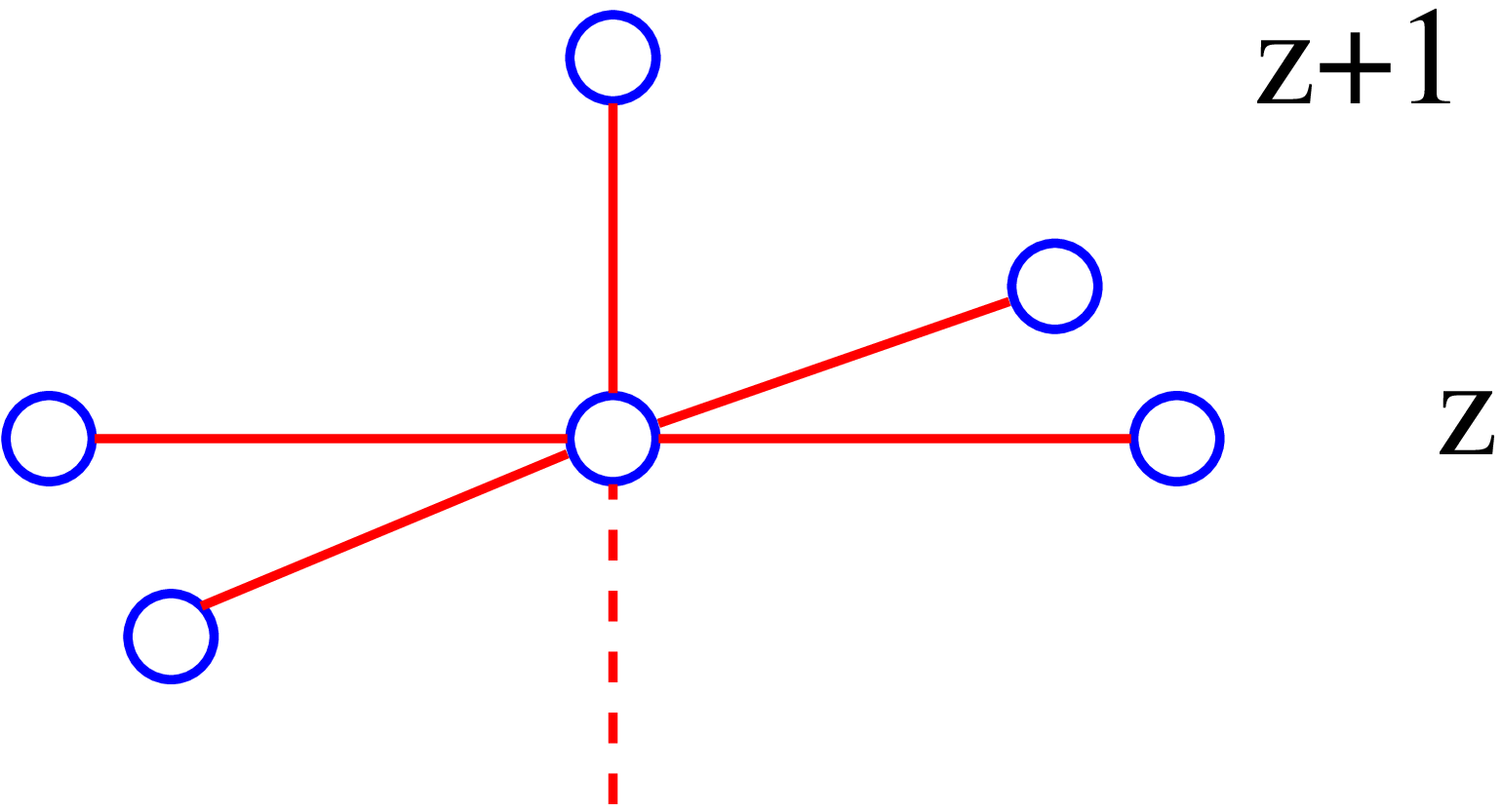}
\includegraphics[scale=0.27]{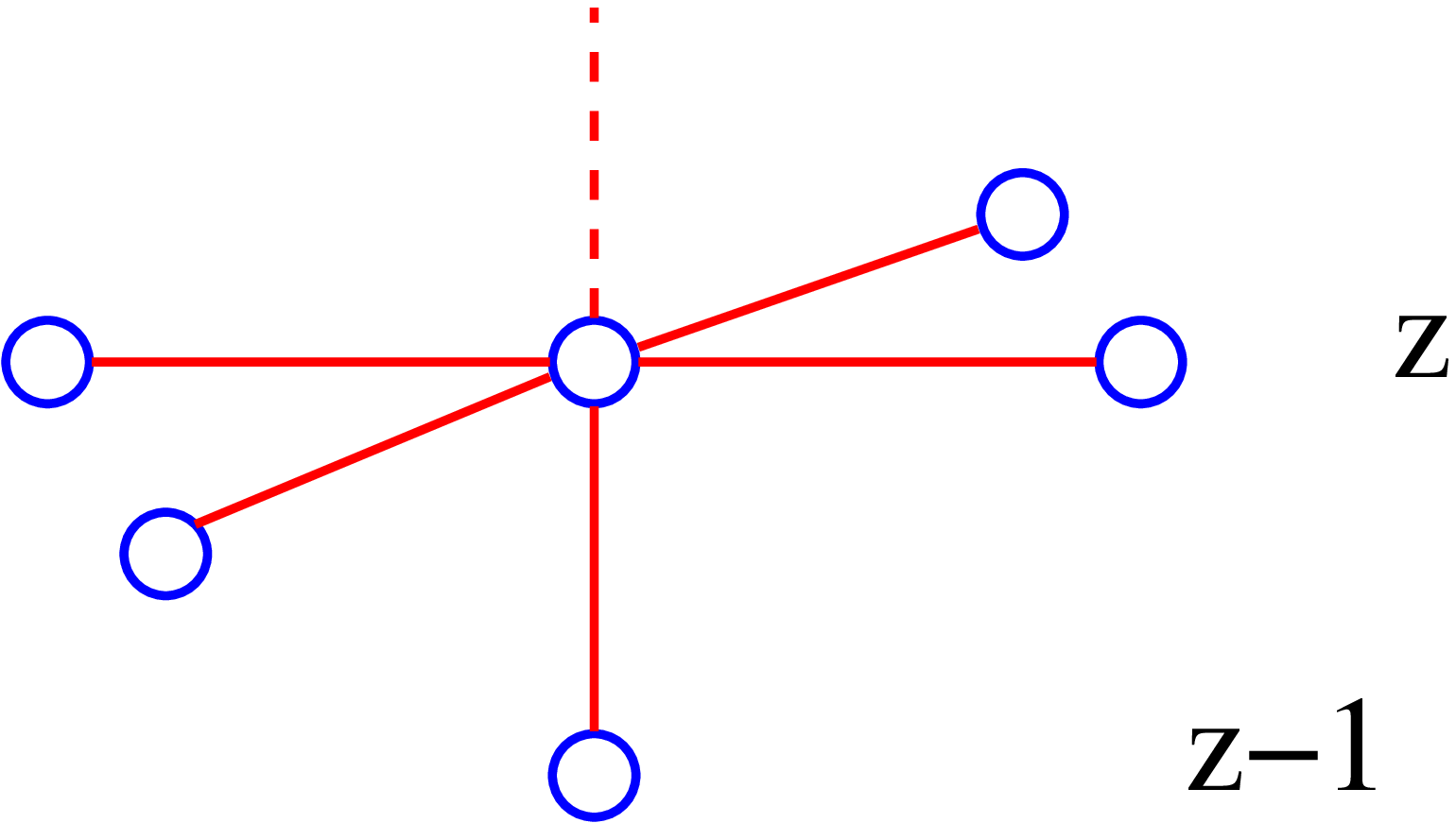}
 \caption{\footnotesize
Three kinds of branches exist here:
1) ``Side'' branch: the root site is connected to
$k-2$  neighbors on its layer, one in the upper
and one in the lower layer;
2) ``Up'' branch: the root site is connected to
$k-1$  neighbors on its layer and one in the upper
layer;
3) ``Down'' branch: the root site is connected to
$k-1$  neighbors on its layer, one in the
lower layer.}
\label{fig:branches}
 \end{center}
\end{figure}

The random graph, Fig.~\ref{Blattice},
keeps into account that the gravity breaks up the symmetry along the $z$
axis. This lattice is made up by $H$ horizontal layers \cite{nota_new}
occupied by hard
spheres (two numbers identify a site of the lattice: The height of the layer,
$z\in\{1,...,H\}$, and the position in the layer, $i$).
Each layer is a random graph of connectivity, $k-1=3$.
Each site in layer at height $z$ is also
connected to its homologous site in $z-1$ and $z+1$
(the total connectivity is thus $k+1$).
The local tree-like structure of the lattice
allows to write down iterative equations \'a la Bethe, where the partition
function of each site is written in terms of the partition functions of the
neighbor sites. We have to introduce the concept of ``branch'': a branch is a
graph where a root site, $i$, has only $k$ neighbors. In the present case
three kinds of branches exist (see Fig.~\ref{fig:branches}):
``up'' (resp. ``down'') branch where the root site has $k-1$ neighbors
on its same layer and one in the upper (resp. lower) layer;
and ``side'' branch where the root has $k-2$ neighbors on its layer,
one in the upper and one in lower layer.


Define $Z_{0,s}^{(i,z)}$ and  $Z_{1,s}^{(i,z)}$ the partition functions
of a ``side'' branch with root site $i$ at height $z$
restricted respectively to configurations
in which the site $i$ is empty or filled by a particle.
$Z_{1,u}^{(i,z)}$ and  $Z_{0,u}^{(i,z)}$ (resp. $\overline{Z}_{0,u}^{(i,z)}$)
are the partition functions of the ``up'' branch restricted respectively to
configurations in which the site $i$ is filled by a particle, or empty
with the upper site filled (resp. empty).  Finally $Z_{1,d}^{(i,z)}$ and
$Z_{0,d}^{(i,z)}$ (resp. $\overline{Z}_{0,d}^{(i,z)}$) are the partition
functions  of the ``down'' branch restricted respectively to
configurations in which the site $i$ is filled by a particle, or empty
with the lower site empty (resp. filled).

The partition function of the branch ending in site $i$ can be recursively
written in terms of the partition functions of the neighbor sites.
Summing over all the possible configurations of the neighbor sites, we obtained
that the
partition function of a ``side'' branch with root site $i$ at height $z$ is:
\begin{eqnarray} \label{recZs}
Z_{0,s}^{(i,z)} & = &
\left[ \prod_{j=1}^{k-2} \left(Z_{0,s}^{(j,z)} + Z_{1,s}^{(j,z)} \right) \right] \\
\nonumber
& \times &
\Big[ Z_{1,u}^{(i,z+1)} \left( Z_{1,d}^{(i,z-1)} + e^{-K} \left( \overline{Z}_{0,d}^{(i,z-1)}
+ Z_{0,d}^{(i,z-1)} \right) \right) \\
\nonumber
& + &
\left(
\overline{Z}_{0,u}^{(i,z+1)} + e^{-K} Z_{0,u}^{(i,z+1)} \right) \\
\nonumber
& \times & \left(
Z_{1,d}^{(i,z-1)} + \overline{Z}_{0,d}^{(i,z-1)} + Z_{0,d}^{(i,z-1)} \right) \Big] \\
\nonumber
Z_{1,s}^{(i,z)} & = & e^{\beta(\mu-mgz)} \left( \prod_{j=1}^{k-2} Z_{0,s}^{(j,z)} \right)
\left( \overline{Z}_{0,d}^{(i,z-1)}+ e^{-K} Z_{0,d}^{(i,z-1)} \right) \\
\nonumber
& \times &
 \left( \overline{Z}_{0,u}^{(i,z+1)} + Z_{0,u}^{(i,z+1)} \right).
\end{eqnarray}
In the same way we can write the recursion relations for the ``up'' branch:
\begin{eqnarray} \label{recZu}
\nonumber
Z_{0,u}^{(i,z)} & = & \left[ \prod_{j=1}^{k-1} \left(Z_{0,s}^{(j,z)} + Z_{1,s}^{(j,z)} \right)
\right]
Z_{1,u}^{(i,z+1)} \\
\overline{Z}_{0,u}^{(i,z)} & = & \left[ \prod_{j=1}^{k-1} \left(Z_{0,s}^{(j,z)} + Z_{1,s}^{(j,z)} \right)
\right] \overline{Z}_{0,u}^{(i,z+1)} \\
\nonumber
Z_{1,u}^{(i,z)} & = & e^{\beta(\mu-mgz)}
\left( \prod_{j=1}^{k-1} Z_{0,s}^{(j,z)} \right)
\left( \overline{Z}_{0,u}^{(i,z+1)} + Z_{0,u}^{(i,z+1)} \right).
\end{eqnarray}
Finally for the ``down'' branch we have:
\begin{eqnarray} \label{recZd}
\nonumber
Z_{0,d}^{(i,z)} & = & \left[ \prod_{j=1}^{k-1} \left(Z_{0,s}^{(j,z)} + Z_{1,s}^{(j,z)} \right)
\right] \left (Z_{0,d}^{(i,z-1)} + \overline{Z}_{0,d}^{(i,z-1)} \right) \\
\overline{Z}_{0,d}^{(i,z)} & =&\left[\prod_{j=1}^{k-1} \left(Z_{0,s}^{(j,z)} + Z_{1,s}^{(j,z)} \right)
\right] Z_{1,d}^{(i, z-1)} \\
\nonumber
Z_{1,d}^{(i,z)} & = & e^{\beta(\mu-mgz)} \left(
\prod_{j=1}^{k-1} Z_{0,s}^{(j,z)} \right) \\
\nonumber
& \times & \left( \overline{Z}_{0,d}^{(i,z-1)}
+ e^{-K} Z_{0,d}^{(i,z-1)} \right).
\end{eqnarray}
In the following we consider the limit $K \to \infty$ in order to take into
account the constraint on the mechanical stability.
It is convenient to introduce five local ``cavity'' fields on each site
$h_s^{(i,z)}$, $h_u^{(i,z)}$, $g_u^{(i,z)}$, $h_d^{(i,z)}$ and $g_d^{(i,z)}$,
defined by the following relations:
$e^{\beta h_s^{(i,z)}} = Z_{1,s}^{(i,z)}/Z_{0,s}^{(i,z)}$;
$e^{\beta h_u^{(i,z)}} = Z_{1,u}^{(i,z)}/\overline{Z}_{0,u}^{(i,z)}$;
$e^{\beta g_u^{(i,z)}} = Z_{0,u}^{(i,z)}/\overline{Z}_{0,u}^{(i,z)}$;
$e^{\beta h_d^{(i,z)}} = Z_{1,d}^{(i,z)}/\overline{Z}_{0,d}^{(i,z)}$;
$e^{\beta g_d^{(i,z)}} = Z_{0,d}^{(i,z)}/\overline{Z}_{0,u}^{(i,z)}$.
In these new variables the recursion relations are more easily written:
\begin{eqnarray} \label{ricorrenza}
\nonumber
e^{\beta h_s^{(i,z)}} & = & e^{\beta(\mu-mgz)}\left[\prod_{j=1}^{k-2}
(1+e^{\beta h_s^{(j,z)}})^{-1}\right]
(1+e^{\beta g_u^{(i,z+1)}})\\
\nonumber
&\times &
[1+e^{\beta h_d^{(i,z-1)}}+e^{\beta g_d^{(i,z-1)}}+e^{\beta h_d^{(i,z-1)}
+\beta h_u^{(i,z+1)}}]^{-1}\\
\nonumber
e^{\beta h_u^{(i,z)}} & = & e^{\beta(\mu-mgz)}
( 1+e^{\beta g_u^{(i,z+1)}} ) \prod_{j=1}^{k-1}
(1+e^{\beta h_s^{(j,z)}})^{-1}
\\
e^{\beta g_u^{(i,z)}} & = & e^{\beta h_u^{(i,z+1)}} \\
\nonumber
e^{\beta h_d^{(i,z)}} & = & e^{\beta(\mu-mgz)}
e^{-\beta h_d^{(i,z-1)}} \prod_{j=1}^{k-1}
(1+e^{\beta h_s^{(j,z)}})^{-1} \\
\nonumber
e^{\beta g_d^{(i,z)}} & = & (1+e^{\beta g_d^{(i,z-1)}})
e^{-\beta h_d^{(i,z-1)}} ~ .
\end{eqnarray}
Note that in the case $k=1$ the problem reduces to a simple
one-dimensional chain: In this case the recursive method is
equivalent to the transfer matrix method and gives the exact solution.

\begin{figure}
 \begin{center}
    \includegraphics[scale=0.33]{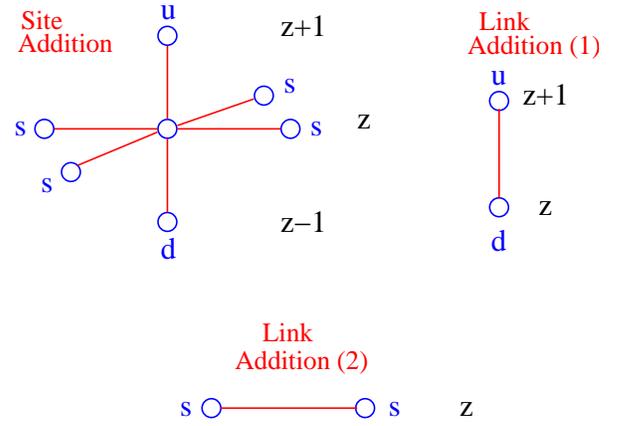}
 \caption{\footnotesize Site addition: a new central site at height $z$
 is connected to
 $k-1$ side branches (s) with the root sites in the same layer,
 to one up (u) branch with the root site in the higher layer
 and to one down (d) branch with the root site
 in the lower layer; Link addition (1): a link
 between a down branch with the root site at height $z$
 and an up branch with the root site in the higher layer is added;
 Link addition (2): a link
  between two side branches with the root site in the same layer is added.
}
\label{shifts}
\end{center}
\end{figure}

From the iterative solution of Eq.s~(\ref{ricorrenza}) it is possible to
compute the system free energy.
Generalizing the procedure followed in \cite{MP} we calculate the free energy
density, $F$, in the thermodynamic limit from the variation of the free energy
going from a random graph with $H$ layers and $N$ sites on each layer to one
with $H$ layers and $N+2$ sites on each layer. In order to do that we define
the following intermediate object:
a random graph with $H$ layers and $N$ sites in each plane such that $2 (k+1)$
sites on each plane are connected only to $k$ neighbors instead of $k+1$.
In particular on each layer 2 sites
are not connected to sites on the higher layer (``down'' branches),
2 sites are not connected to sites on the lower layer (``up''
branches) and the other $2(k-1)$ are connected
only with $k-2$ sites in the plane instead of $k-1$ (``side'' branches).
From this intermediate object a random graph with $H$ layers and $N+2$ sites
on each layer (all connected to $k+1$ sites)
can be obtained adding $2$ new sites to each plane and connecting
each of the new sites with $k-1$ side branches on their respective planes,
one up branch in the upper layer and one down branch in the lower layer
(see Fig.~\ref{shifts}).  This operation is called ``site addition''.
A random graph with $H$ layers and $N$  sites on each layer (all connected to
$k+1$ sites) is instead obtained from the intermediate object
adding for each layer $2$ links
between the up branches at height $z$ and the down branches at height $z-1$,
and $(k-1)$ links between the sides branches on each layer
(see Fig.~\ref{shifts}).  This operation, which allows to saturate all the
missing links, is called ``link addition''.

Therefore the variation of the free energy when going from $NH$ to
$(N+2)H$ sites (i.e. a random graph with two sites more
on each layer)
is related to the free energy shifts (see Fig.~\ref{shifts})
for a site addition ($\Delta F_{site}^{(z)}$) and for
two different kinds of
link addition ($\Delta F_{link, 1}^{(z)}$ and $\Delta F_{link, 2}^{(z)}$):
\begin{eqnarray}
\nonumber
F_{N+2} - F_{N} & = & 2 \sum_{z=1}^H \Delta F_{site}^{(z)} - (k-1)
\sum_{z=1}^H \Delta F_{link,2}^{(z)} \\
\nonumber
&-& 2 \sum_{z=1}^{H-1} \Delta F_{link,1}^{(z)},
\end{eqnarray}
where $F_{N+2} - F_{N}$ is obtained as $(F_{N+2}-F_0) - (F_{N}-F_0)$, and $F_0$
is the free energy of the intermediate object described above.
We assume that in the thermodynamic limit the free energy is linear in $N$. The
free energy density is then:
\begin{equation} \label{free_energy}
F = \sum_{z=1}^H \Delta F_{site}^{(z)} - \frac{(k-1)}{2} \sum_{z=1}^H
\Delta F_{link,2}^{(z)}
- \sum_{z=1}^{H-1} \Delta F_{link,1}^{(z)}.
\end{equation}
In terms of the local fields the free energy shifts due to
the addition of a site $i$ at height $z$ reads:
\begin{eqnarray} \label{site}
&&e^{-\beta \Delta F^{(i,z)}_{site}}  =  \left[ \prod_{j=1}^{k-1} \left( 1+ e^{\beta h_s^{(i,z)}} \right)
\right] \\
\nonumber
&& \times
\left(1+e^{\beta h_d^{(i,z-1)}} +e^{\beta g_d^{(i,z-1)}} +e^{\beta h_d^{(i,z-1)}
} e^{\beta h_u^{(i,z+1)}} \right) \\
\nonumber
&& + e^{\beta(\mu-mgz)} \left( 1+e^{\beta g_u^{(i,z+1)}} \right).
\end{eqnarray}
The free energy shift due to a link addition between a down branch
with the root site at height $z$ and an up branch with the root site
at height $z+1$ is given by:
\begin{eqnarray} \label{link1}
e^{-\beta \Delta F^{(i,z|z+1)}_{link,1}} & = & 1+e^{\beta g_d^{(i,z)}}+
e^{\beta h_u^{(i, z+1)}}\\
\nonumber
&+&e^{\beta h_d^{(i,z)}}(1+e^{\beta g_u^{(i,z+1)}}).
\end{eqnarray}
Finally the free energy shift due to a link addition between two side branches
with root sites $i$ and $j$ at height $z$ is:
\begin{equation} \label{link2}
e^{-\beta \Delta F^{(i|j,z)}_{link,2}} = 1 + e^{\beta h_s^{(i,z)}}+e^{\beta
h_s^{(j,z)}}.
\end{equation}
In order to compute the free energy of the system we have to compute
the mean values of the free energy shifts for all the sites at a given height
and for all the possible realization of the lattice.
In the following these mean values will be computed in three different
cases: 1) A fluid-like homogeneous phase; 2) A crystalline-like solution
characterized by the breakdown of the horizontal translational invariance;
3) A glassy phase by a 1-step Replica Symmetry Breaking.

The fluid-like solution is obtained by setting
that the local fields on each layer are the same for all sites
of the layer ($\{\mathbf{h}^{(i,z)}\} = \{\mathbf{h}^{(z)}\}
\,\,\,\, \forall i$).  In this case
Eq.s~(\ref{ricorrenza}) become $5 H -1$ algebraic coupled equations
and they are easily solved finding the fixed points.
This homogeneous (Replica Symmetric) solution is
characterized by horizontal translational invariance and
is found to be stable for high values of the configurational temperature,
$T_{conf}$, or for low values of the number of grains per
unit surface, $N_s$.
In this case the free energy
is easily computed from Eq.s~(\ref{site}), (\ref{link2}) and (\ref{link1}),
since in this case all the quantities are site independent.
From the free energy, $F$, we derive the density profile, $\sigma (z)\equiv
\langle n_i(z)\rangle$:
\begin{equation}
\sigma (z) = \frac{e^{\beta(\mu-mgz)} \left( 1+e^{\beta g_u^{(z+1)}} \right)}
{e^{-\beta \Delta F^{(z)}_{site}}},
\end{equation}
the number of particles per unity of surface, $N_s\equiv\sum_{z=1}^H \sigma(z)$,
and the gravitational energy density,
$E \equiv \sum_{z=1}^H mgz \sigma(z)$.
From the relation $F= E - TS -\mu N_s$, we also calculate
the entropy per lattice site,
$S = -\beta F -\beta \mu N_s + \beta E$.

In the crystalline (Replica Symmetric) solution
the local fields are different on different sites
(breakdown of translational invariance), but
do not fluctuate from site to site.
This is achieved by the introducing two sub-lattices, \emph{a} and \emph{b}, and
different local fields on each lattice. The merging is done
taking into account the structure of the crystalline phase.
In our case, each site of the sub-lattice \emph{a} (resp. \emph{b})
is connected with $k+1$ sites of the sub-lattice \emph{b} (resp. \emph{a}).
The crystal periodicity is thus two lattice spacings.
Schematically Eq.s~(\ref{ricorrenza}) for each layer become:
\begin{displaymath}
\left \{
\begin{array}{l}
\{ \bf{h}_a \} = \mathbf{f} (\beta, \mu, \{ \bf{h}_b \}) \\
\{ \bf{h}_b \}  =  \mathbf{f} (\beta, \mu, \{ \bf{h}_a \})
\end{array}
\right.
\end{displaymath}
where $\{ \bf{h}_a \}$ and $\{ \bf{h}_b \}$ are the sets of all local fields
respectively on the two sub-lattices.
This is a system of $2(5 H -1)$ algebraic coupled equations.
The free energy is computed from the fixed points of these equations.
For a given $N_s$, by lowering $T_{conf}$, a phase
transition from the fluid to the crystal is found
at $T_m$ (see Fig.~\ref{MFPD}).

The fluid phase still exist below $T_m$ as a metastable phase
corresponding to a supercooled fluid when crystallization
is avoided.
Nevertheless, the entropy per site predicted by the fluid
solution becomes negative when the temperature is lowered,
or the packing fraction is increased.
Hence, this
solution is not appropriate to describe the high $N_s$
or low $T_{conf}$ region. A solution characterized
by the presence of a large number of local minima of the
free energy is found in this region.
In this case the local fields may fluctuate.
To describe this situation we have to introduce three
probability distributions on each layer
$\mathcal{P}_{i,z}^{u} (h_u, g_u)$, $\mathcal{P}_{i,z}^{s} (h_s)$
and $\mathcal{P}_{i,z}^d (h_d, g_d)$ defined as the probability of finding the
fields $h_u^{(i,z)}$ and $g_u^{(i,z)}$ (or respectively $h_s^{(i,z)}$, or
$h_d^{(i,z)}$ and $g_d^{(i,z)}$) on site $i$ at height $z$ equal to
$h_u$ and $g_u$ (or respectively to $h_s$, or to $h_d$ and $g_d$).
Since the glassy phase is expected to be translational invariant, we work
in the factorized case in which the probability distributions at a given height
are equal for all the sites of the layer ($\mathcal{P}_{i,z}^{u,s,d} \equiv
\mathcal{P}_{z}^{u,s,d}$).

Within the one-step Replica Symmetry Breaking ansatz of the cavity method
(see appendix \ref{app2}) the recursion relations for the fields are replaced
by self consistent integral equations for the probability
distribution of the fields.
For the ``up'' merging the self consistent integral equation reads:
\begin{eqnarray} \label{pup}
\mathcal{P}_{z}^{u} (h_u^z, g_u^z) & = & \mathcal{C}_1 \int \prod_{j=1}^{k-1}
\left[ d h_{s}^{(j,z)} \mathcal{P}_{z}^{s} (h_s^{(j,z)}) \right] \\
\nonumber
& \times &
\left [d h_{u}^{(i,z+1)} d g_{u}^{(i,z+1)} \mathcal{P}_{z+1}^{u} (h_u^{(i,z+1)}, g_u^{(i,z+1)})
\right]\\
\nonumber
&\times &
\delta (h_{u}^{z} - h_u^{(i,z)}) \delta (g_{u}^{z} - g_u^{(i,z)})
e^{-\beta m \Delta F_{up}^{(i,z)}},
\end{eqnarray}
where $\mathcal{C}_1$ is a constant
insuring the normalization of $\mathcal{P}_{z}^{u}$,
$h_u^{(i,z)}$ and $g_u^{(i,z)}$ are the local fields defined by Eq.s
~(\ref{ricorrenza}), $m \in [0,1]$ is the usual 1RSB parameter to be
obtained by the maximization of the free energy with respect to it, and
$\Delta F_{up}^z$ is the free energy shift in the ``up'' merging process.
This quantity is computed by using that the addition of a
site $i$ at a certain height $z$ (``site addition'') is
the result of an ``up'' merging process, which bring to a new
``up'' branch with root site $i$, plus a link addition between this
branch and a down branch at height $z-1$:
\begin{equation} \label{deltafup}
\Delta F^{(i,z)}_{site} = \Delta F^{(i,z)}_{up} + \Delta F^{(i,z-1|z)}_{link,1}.
\end{equation}
From this equation we obtain that:
\begin{equation}
e^{- \beta \Delta F^{(i,z)}_{up}} =
\frac{\overline{Z}_{0,u}^{(i,z)}}{\overline{Z}_{0,u}^{(i,z+1)}
\prod_{j=1}^{k-1}
Z_{0,s}^{(j,z)}}.
\label{A12}
\end{equation}
From Eq.s~(\ref{recZs}-\ref{recZd}) the free energy
shift $\Delta F^{(i,z)}_{up}$ has a simple expression in terms of the
local fields.

In the same way we can determine the self consistency
equations for the other two kinds
of merging:
\begin{eqnarray} \label{pside}
\mathcal{P}_{z}^{s} (h_s^z) & = & \mathcal{C}_2 \int \prod_{j=1}^{k-2}
\left[ d h_{s}^{(j,z)} \mathcal{P}_{z}^{s} (h_s^{(j,z)}) \right] \\
\nonumber
& \times &
\left [d h_{d}^{(i,z-1)} d g_{d}^{(i,z-1)} \mathcal{P}_{z-1}^{d} (h_d^{(i,z-1)}, g_d^{(i,z-1)})
\right] \\
\nonumber
& \times &
\left [d h_{u}^{(i,z+1)} d g_{u}^{(i,z+1)} \mathcal{P}_{z+1}^{u} (h_u^{(i,z+1)}, g_u^{(i,z+1)})
\right] \\
\nonumber
& \times &
\delta (h_{s}^{z} - h_s^{(i,z)})
e^{-\beta m \Delta F_{side}^{(i,z)}},
\end{eqnarray}
\noindent
and
\begin{eqnarray} \label{pdown}
\mathcal{P}_{z}^{d} (h_d^z, g_d^z) & = & \mathcal{C}_3 \int \prod_{j=1}^{k-1}
\left[ d h_{s}^{(j,z)} \mathcal{P}_{z}^{s} (h_s^{(j,z)}) \right] \\
\nonumber
& \times &
\left [d h_{d}^{(i,z-1)} d g_{d}^{(i,z-1)} \mathcal{P}_{z-1}^{d} (h_d^{(i,z-1)}, g_d^{(i,z-1)})
\right]\\
\nonumber
& \times &
\, \delta (h_{d}^{z} - h_d^{(i,z)}) \delta (g_{d}^{z} - g_d^{(i,z)})
e^{-\beta m \Delta F_{down}^{(i,z)}}.
\end{eqnarray}
For the ``side'' and the ``down'' merging one has that:
\begin{equation}
\Delta F^{(i,z)}_{site} = \Delta F^{(i,z)}_{side} +
\Delta F^{(i|j,z)}_{link,2},
\end{equation}
and
\begin{equation}
\Delta F^{(i,z)}_{site} = \Delta F^{(i,z)}_{down} + \Delta F^{(i,z|z+1)}_{link,1}.
\end{equation}
This yields to:
\begin{equation}
e^{- \beta \Delta F^{(i,z)}_{side}} =
\frac{Z_{0,s}^{(i,z)}}{\overline{Z}_{0,u}^{(i,z+1)}
\overline{Z}_{0,d}^{(i,z-1)}
\prod_{j=1}^{k-2} Z_{0,s}^{(j,z)}},
\end{equation}
and
\begin{equation}
e^{- \beta \Delta F^{(i,z)}_{down}} =
\frac{\overline{Z}_{0,d}^{(i,z)}}{\overline{Z}_{0,d}^{(i,z+1)} \prod_{j=1}^{k-1}
Z_{0,s}^{(j,z)}}.
\end{equation}
For any value of $\beta$, $\mu$ and $m$ we solve
Eq.s~(\ref{pup}), (\ref{pside}) and (\ref{pdown}) iteratively,
discretizing the probability distributions until the whole procedure converged.

From the probability distributions we compute the free energy density
of the system: according to Eq.~(\ref{free_energy}) we have to find the average
values of the free energy shifts due to link and site additions.
The free energy shift due to site addition is given by:
\begin{eqnarray}
\langle e^{-\beta m \Delta F_{site}{(z)}}\rangle & = & \int \prod_{j=1}^{k-1}
\left[ d h_{s}^{(j,z)} \mathcal{P}_{z}^{s} (h_s^{(j,z)}) \right] \\
\nonumber
& \times &
\left [d h_{d}^{(i,z-1)} d g_{d}^{(i,z-1)} \mathcal{P}_{z-1}^{d}
(h_d^{(i,z-1)}, g_d^{(i,z-1)})
\right] \\
\nonumber
&\times&
\left[ d h_{u}^{(i,z+1)} d g_{u}^{(i,z+1)} \mathcal{P}_{z+1}^{u}
(h_u^{(i,z+1)}, g_u^{(i,z+1)})
\right] \\
\nonumber
& \times& e^{-\beta m \Delta F_{site}{(i,z)}}.
\end{eqnarray}
For the first kind of link addition we have:
\begin{eqnarray}
\nonumber
&&\langle e^{-\beta m \Delta F_{link,1}^{(z)}}\rangle  =  
\\
\nonumber
&&\int
\left [d h_{d}^{(i,z-1)} d g_{d}^{(i,z-1)}
\mathcal{P}_{z-1}^{d} (h_d^{(i,z-1)}, g_d^{(i,z-1)})
\right]
\\
\nonumber
&&\times
\left[
d h_{u}^{(i,z+1)} d g_{u}^{(i,z+1)} \mathcal{P}_{z+1}^{u}
(h_u^{(i,z+1)}, g_u^{(i,z+1)})
\right] \\
&&\times e^{-\beta m \Delta F_{link,1}^{(i,z-1|z)}}.
\end{eqnarray}
Finally for the second kind of link addition we find:
\begin{equation}
\langle e^{-\beta m \Delta F_{link,2}^{(z)}} \rangle= \int \prod_{j=1}^{2}
\left[ d h_{s}^{(j,z)} \mathcal{P}_{z}^{s} (h_s^{(j,z)}) \right]
e^{-\beta m \Delta F_{link,2}^{(i|j,z)}}.
\end{equation}
In the previous relations $\Delta F_{site}{(i,z)}$,
$\Delta F_{link,1}^{(i,z-1|z)}$ and $\Delta F_{link,2}^{(i|j,z)}$
are function of the local fields
according to Eq.~(\ref{site}), (\ref{link1}) and (\ref{link2}).

The total free energy density of the system is, according to Eq.
~(\ref{free_energy}):
\begin{eqnarray}
F[m] &=& -\,\,\frac{1}{\beta m} \Bigg[
\sum_{z=1}^H \log e^{- \beta m \Delta F_{site}^{(z)}} \\
\nonumber
&-&
\sum_{z=1}^H \frac{(k-1)}{2} \log e^{- \beta m \Delta F_{link,2}^{(z)}} \\
\nonumber
&-& \sum_{z=1}^{H-1} \log e^{-\beta m \Delta F_{link,1}^{(z)}} \Bigg].
\end{eqnarray}
The parameter $m$ is fixed by the maximization of the free energy with respect
to it. The justification for that is in the replica method, since
$m$ turns out to be the breakpoint in Parisi's order parameter function
at the 1-step RSB level. For a spin glass it has been rigorously proved that
in the limit $k \to \infty$, $F[m]$ is a lower bound to the correct free energy,
so it is natural to find the preferred value of $m$ by the maximization of
$F[m]$.

\section{Self-consistency equations in the cavity method}
\label{app2}
In this appendix we show how to obtain the self-consistency integral
equations (\ref{pup}), (\ref{pside}) and (\ref{pdown}) using the
cavity method in the 1-step RSB ansatz \cite{MP,Biroli}.
The region at high packing fraction (or at low configurational
temperature) is characterized by the existence of many pure states.
Let $\mathcal{N} (F)$ the number of pure states for a given value of the
free energy of the system.
The function $\Sigma (F) = \log \mathcal{N} (F)$ is called
complexity. We assume that within one pure states $\alpha$ the local fields
$h_{u,\alpha}^{(i,z)}$, $g_{u,\alpha}^{(i,z)}$, $h_{s,\alpha}^{(i,z)}$,
$h_{d,\alpha}^{(i,z)}$ and $g_{d,\alpha}^{(i,z)}$ on different
cavity sites are uncorrelated.
Therefore Eq.s~(\ref{ricorrenza}) continue to hold in any given pure state.
In this case we have to make a statistical description
of the solutions of Eq.s~(\ref{ricorrenza}) in the different pure states,
taking into account the number of pure states for a given value of the free
energy.

Let us consider, for example, the ``up'' merging of $k$ cavity sites
in a site $i$ at height $z$.  As said before, in each pure state $\alpha$
the local fields in the $k$ cavity
sites are not correlated. Nevertheless in each pure state $\alpha$
the local fields, $(h_{u,\alpha}^{(i,z)}, g_{u,\alpha}^{(i,z)})$,  and
the free energy shift, $\Delta F_{u,\alpha}^{(i,z)}$, due to the merging
are correlated,
since they are both functions of the local fields in
the neighbor sites in the state $\alpha$, according to Eq.s~(\ref{ricorrenza})
and (\ref{deltafup}). Let us define $\mathcal{S}_z (h_{u}^{z}, g_{u}^{z},
\Delta F_{u}^{z})$ as the probability distribution of finding the fields
$(h_{u,\alpha}^{(i,z)}, g_{u,\alpha}^{(i,z)})$ and the free energy shift $\Delta
F_{u}^{z}$ after an up merging at height $z$.
Because of the recursion relations of Eq.s
~(\ref{ricorrenza}) and (\ref{deltafup}),
this distribution probability has to verify the
following iteration relation:
\begin{eqnarray}
\mathcal{S}_z (h_{u}^{z}, g_{u}^{z}, \Delta F_{u}^{z}) & = & \int \prod_{j=1}^{k-1}
\left[ d h_{s}^{(j,z)} \mathcal{P}_{z}^{s} (h_s^{(j,z)}) \right] \\
\nonumber
&\times&
\Big[ d h_{u}^{(i,z+1)} d g_{u}^{(i,z+1)} \\
\nonumber
&\times& \mathcal{P}_{z+1}^{u} (h_u^{(i,z+1)}, g_u^{(i,z+1)})
\Big] \\
\nonumber
&\times&
\delta (h_{u}^{z} - h_u^{(i,z)}) \delta (g_{u}^{z} - g_u^{(i,z)}) \\
\nonumber
&\times&
\delta (\Delta F_{u}^{z} - \Delta F_{u}^{(i,z)}).
\end{eqnarray}
In order to determine the probability distribution for the local
fields self-consistently
we have to make the integration over all possible free energy shifts:
\begin{eqnarray}
\nonumber
\mathcal{P}_{z}^{u} (h_u^z, g_u^z) & = & \int d(\Delta F_u^z)
\mathcal{S}_z (h_{u}^{z}, g_{u}^{z}, \Delta F_{u}^{z}) \mathcal{N} (F - \Delta F) \\
\nonumber
& = &  \int d(\Delta F_u^z)
\mathcal{S}_z (h_{u}^{z}, g_{u}^{z}, \Delta F_{u}^{z})
\e^{\Sigma \left(F - \Delta F_u^z\right)}.
\end{eqnarray}
Since we are interested only in the local minima with the lowest
free energies, we
expand the exponent to the first order in $\Delta F_u^z$:
\begin{equation}
\mathcal{P}_{z}^{u} (h_u^z, g_u^z) = \mathcal{C}_1 \int  d(\Delta F_u^z)
\mathcal{S}_z (h_{u}^{z}, g_{u}^{z}, \Delta F_{u}^{z}) \exp (-\beta m \Delta F_u^z),
\end{equation}
\noindent
where the parameter $m \in [0,1]$ is:
\begin{equation}
m = \frac{1}{\beta} \frac{\partial \Sigma}{\partial F},
\end{equation}
\noindent
and $\mathcal{C}_1$
is a normalization constant. Actually the first order expansion
means that the density of pure states for a given value of the free energy is
$\mathcal{N} \simeq \exp (m(F - F_{ref}))$, where $F_{ref}$ is a reference
free energy whose value is completely irrelevant.
This form of the density of states
is the same found in the 1-step RSB formulation.

By integrating over $\Delta F_u^z$, the delta function selects only the
right value of the free energy shift given in Eq.~(\ref{A12}). We thus have:
\begin{eqnarray}
\mathcal{P}_{z}^{u} (h_u^z, g_u^z) & = & \mathcal{C}_1 \int \prod_{j=1}^{k-1}
\left[ d h_{s}^{(j,z)} \mathcal{P}_{z}^{s} (h_s^{(j,z)}) \right] \\
\nonumber
&\times&
\left [d h_{u}^{(i,z+1)} d g_{u}^{(i,z+1)} \mathcal{P}_{z+1}^{u} (h_u^{(i,z+1)}, g_u^{(i,z+1)})
\right]\\
\nonumber
&\times &
\delta (h_{u}^{z} - h_u^{(i,z)}) \delta (g_{u}^{z} - g_u^{(i,z)})
\e^{-\beta m \Delta F_u^z}.
\end{eqnarray}
We have thus obtained the self consistency Eq.~(\ref{pup}).
In the same way it is possible to obtain the equations for the ``side''
and the ``down'' merging.


\begin{thebibliography}{100}
\bibitem{notadens} In the case of uniform density profile, i.e.  
$\sigma (z)=const.$, we have $\sigma (z)=\Phi$ (where
$\Phi \equiv N/ L^2(2\langle z \rangle-1)$) below the maximum height and zero 
above.
\bibitem{notapart}
We add a fully empty layer in $z=-1$ and a fully occupied layer in $z=0$ in
order to simulate the box bottom, and
a fully empty layer in $z=L+1$ to allow the particles to occupy the top of
the box without violating the hard core interaction.
\bibitem{notaMCT} The $\beta$-correlator predicted by the MCT is given by
$\Phi(t)=f_c+ g(t/t_\sigma)$, where $g(t/t_\sigma)\propto(t/t_\sigma)^{-a}$
for $t_0\ll t\ll t_\sigma$, and
$g(t/t_\sigma)\propto-(t/t_\sigma)^b$ for $t_\sigma\ll t\ll\tau_\alpha$.
\bibitem{nota_max} Interestingly this anomalous behavior
seems to occur around the crossover temperature $T_D$
previously calculated. The origin of this behavior, also observed in molecular
dynamics simulations of a usual glass former \cite{glotzer},
is still unclear.
\bibitem{nota_new} We add an auxiliary layer at height $H+1$ where all sites
are empty in order to allow the particles to occupy the top of the box without
violating the hard core interaction. For the first two layers, $z=1,~ 2$,
we assume that the particles are
always mechanically stable in order to simulate the presence of the bottom of
the box.
\end{thebibliography}
\end{document}